\documentclass[twocolumn,final]{IEEEtran}
\usepackage[final]{graphicx}
\usepackage[final]{graphicx}
\usepackage{amsmath,mathrsfs,amssymb, mathtools,multicol,multirow}
\newcommand{\bc}{\begin{center}}
\newcommand{\ec}{\end{center}}
\newcommand{\bdm}{\begin{displaymath}}
\newcommand{\edm}{\end{displaymath}}
\newcommand{\beq}{\begin{equation}}
\newcommand{\eeq}{\end{equation}}
\newcommand{\bfl}{\begin{flushleft}}
\newcommand{\efl}{\end{flushleft}}
\newcommand{\bt}{\begin{tabbing}}
\newcommand{\et}{\end{tabbing}}

\usepackage{here}
\numberwithin{equation}{section} \allowdisplaybreaks

\begin{document}

\title{Understanding Distal Transcriptional Regulation from Sequence Motif, Network Inference and Interactome Perspectives}

\author{ Arvind~Rao, Alfred~O.~Hero~III, David~J.~States, James~Douglas~Engel
\thanks{Arvind~Rao and Alfred~O.~Hero,~III are with the Departments of Electrical
Engineering and Computer Science, and Bioinformatics at the
University of Michigan, Ann Arbor, MI-48109, email: [ ukarvind,
hero] @umich.edu}
\thanks{James~Douglas~Engel is with the Department of Cell and Developmental Biology at the University of Michigan, Ann Arbor, MI-48109.}
\thanks{David~J.~States is with the Departments of Bioinformatics and Human Genetics at the University of Michigan, Ann Arbor, MI-48109.}
\thanks{A part of this work has been presented at the Computational Systems Bioinformatics (CSB) 2007 conference.}
}

\maketitle

\begin{abstract}
Gene regulation in higher eukaryotes involves a complex interplay between the gene proximal promoter and distal genomic elements (such as enhancers) which work in concert to drive spatio-temporal expression. The experimental characterization of gene regulatory elements is a very complex and resource-intensive process. One of the major goals in computational biology is the \textit{in-silico} annotation of previously uncharacterized elements using results from the subset of known, annotated, regulatory elements.

The recent results of the ENCODE project presented  in-depth experimental analysis of such functional (regulatory) non-coding elements for $1\%$ of the human genome. This dataset enables the principled association of experimental results with true regulatory activity from \textit{in-vitro} or \textit{in-vivo} studies. It is hoped that the results obtained on this subset can be scaled to the rest of the genome. This is an extremely important effort which will enable faster dissection of other functional elements in key biological processes such as disease progression and organ development.  
The computational annotation of these hitherto uncharacterized regions would require an identification of features that have good predictive value for regulatory behavior.

Though the exact mechanism of gene regulation is not completely known, several data-driven experimental models have been hypothesized to understand transcription, pointing to sequence, expression, transcription factor (TF) and their interactome level attributes, at  the biochemical and biophysical levels. This has largely been possible due to the advent of new techniques in functional genomics, such as TF chromatin immunoprecipitation (ChIP), RNA interference, microarray expression profiling,  yeast-2-hybrid ($Y2H$) screens and chromosome conformation capture studies. However, these features are yet to be meaningfully integrated for understanding transcriptional regulatory mechanisms computationally. It is believed that such data-driven computational models can be extremely useful to the discovery of new regulatory elements of desired function and specificity.

In this work, we study transcriptional regulation as a problem in heterogeneous data integration, across sequence, expression and interactome level attributes. Using the example of the \textit{Gata2} gene and  its recently discovered urogenital enhancers \cite{Khandekar2004} as a case study, we examine the predictive value of  various high throughput functional genomic assays in characterizing these enhancers and their regulatory role. Observing results from the application of modern statistical learning methodologies for each of these data modalities, we propose a set of attributes that are most discriminatory in the localization and behavior of these enhancers.
\end{abstract}


\section{Introduction}\label{introduction}
\begin{figure}[!h!t!b]
\centerline{\includegraphics[width=3.9in,height=1.5in]{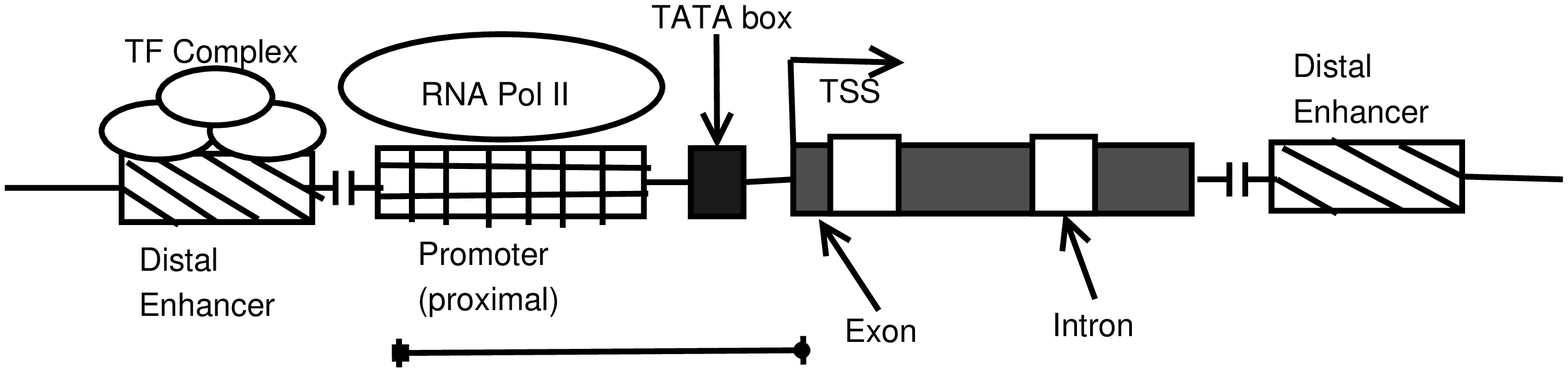}}
\caption{Schematic of Transcriptional Regulation. Sequence motifs at the promoter and the distal regulatory elements together confer specificity of gene expression via TF binding.}\label{fig:transcription1}
\end{figure}

Understanding the mechanisms underlying regulation of tissue-specific gene expression remains a challenging question. While all mature cells in the body have a complete copy of the human
genome, each cell type only expresses those genes it needs to carry out its assigned task. This includes genes required for basic cellular maintenance (often called ``house-keeping genes") and those genes whose function is specific to the particular tissue type that the cell belongs to. Gene expression by way of transcription is the process of generation of messenger RNA (mRNA) from the DNA template representing the gene. It is the intermediate step before the generation of functional protein from messenger RNA. During gene expression, transcription factor (TF) proteins are recruited at the proximal promoter of the gene as well as at sequence elements (enhancers/silencers) which can lie several hundreds of kilobases from the gene's transcriptional start site (Figs. \ref{fig:transcription1} and \ref{fig:transcription2}).

It is hypothesized that the collective set of transcription factors that drive (regulate) expression of a target gene are cell, context and tissue dependent (\cite{EnhancerBrowser}, \cite{EnhancerBrowser2}). Some of these TFs are recruited at proximal regions such as the promoter of the gene, while others are recruited at more distal regions, such as enhancers. There are several (hypothesized) mechanisms for promoter-enhancer interaction through protein interactions between TFs recruited at these elements during formation of the transcriptional complex \cite{looping_scan_track}. A commonly accepted mechanism of distal interaction, during regulation, is looping (\cite{3C/4C}, \cite{looping_2006}, \cite{Mecp2}), shown in Fig. \ref{fig:transcription2}.

To understand the role of various genomic elements in governing gene regulation, functional genomics has played an enabling role in providing heterogeneous data sources and experimental approaches to discern distal interactions during transcription. Each of these experiments have aimed to resolve different aspects (features) of transcriptional regulation focussing on TF binding, promoter modeling and epigenetic preferences for tissue-specific expression in some genomic regulatory elements (\cite{ENCODE},  \cite{chromatin-ChIP}, \cite{Sanger-Histone}, \cite{Segal2006}, \cite{Mecp2}).  Additionally, some studies have demonstrated that these data sets along with principled statistical metrics can be used to derive such features computationally, with a view to asking questions relevant to the biology of transcriptional regulation (\cite{chromatin-ChIP}, \cite{Segal2006}, \cite{Meyer2004}, \cite{Silver2007}).

There have been several principled yet scattered studies characterizing the role of functional regulatory elements for certain genes (such as \textit{Mecp2}, \textit{Shh}, \textit{Gata2}, \textit{Gata3}) in various organisms (\cite{Mecp2}, \cite{Shh}, \cite{Khandekar2004}, \cite{CNSGata3}, \cite{Gata3KE}, \cite{Fgf_enh}). These reveal an inherent spatio-temporal context of gene expression and regulation. However, there is a need for a unified set of principles, spanning various genomic attributes, that could account for the behavior of these tissue-specific and gene-specific enhancers.
We note that there are promoter-independent enhancers too, and their computational study has been far more principled (\cite{EnhancerBrowser}, \cite{Enhancer_Prediction}); however, their study is outside the scope of this study where we focus on gene-specificity in addition to tissue-specificity.

The results of the ENCODE project (\emph{http://encode.nih.gov/}), (\cite{ENCODE}, \cite{Sanger-Histone}) on $1\%$ of the human genome has established some very interesting results about the nature of transcriptional regulation at the genome scale. Particularly, they report the use of several experimental techniques (Histone ChIP on chip, DNASE1 hypersensitivity assays) etc analyzing transcribed regions as well as their regulatory regions, genome-wide. A large scale computational effort is developing alongside to ``learn" features of such regulatory elements and use of these features for predicting other control elements for genes outside the ENCODE regions, thereby accomplishing a genome-wide annotation. Considering that over  $98\%$ of the genome is non-coding, this effort is going to parallel the previous project in gene-annotation at the genome scale in effort and importance. Adding to this complexity is the fact that the same non-coding element can potentially regulate the expression of genes in a spatio-temporal manner, activating different genes at different times in different tissues, and from arbitrarily large distances from the gene. Thus there is a need for the principled ``reverse-engineering" of the architectures of these regulatory elements, using features at the sequence, expression and interactome level.

Understanding the mechanism of transcriptional regulation thus entails several aspects:
\begin{enumerate}
\item
Do regulatory regions like promoters and enhancers have any interesting \emph{sequence properties} depending on their tissue-specificity of gene expression? These properties can  be examined based on their individual sequences or their epigenetic preferences. A common technique of analysis is the identification of tissue-specific motif-signatures (\cite{Fraenkel2006}, \cite{Kreiman2004}) for such elements.
\item
To reduce the vast number of false positives that arise from sequence approaches alone, we appeal to a mechanistic insight from biology. For long-range transcriptional regulation to be enabled, there has to be an enhancer-promoter interaction during formation of the tissue-specific, gene-specific transcriptional machinery. Literature suggests that such interaction is mediated by protein-protein interactions between promoter TFs and enhancer TFs after looping along the chromosomal length (\cite{Mecp2}, \cite{prom_enh_ppi1}, \cite{prom_enh_ppi2}, \cite{3C/4C}). 
This insight (Fig. \ref{fig:transcription2}) poses two further questions:
\begin{itemize}
\item
Which TFs bind the promoter and the putative enhancer?
\item
Do the resultant interactions between enhancer and promoter TFs have any special characteristic that discriminate functional non-coding regulatory regions from non-functional ones?
\end{itemize}
\end{enumerate}

As a case study to answer some of these questions, we examine the regulation of \textit{Gata2} regulation in the developing kidney. \textit{Gata2} is a gene belonging to the GATA family of transcription factors (\textit{GATA1-6}), and binds the consensus -WGATAR- motif on DNA \cite{Gata2}. It is located on mouse chromosome $6$, and plays an important role in mammalian hematopoiesis, nephrogenesis and CNS development, with important phenotypic consequences. The study of long-range regulatory elements that effect \textit{Gata2} expression has been on for several years now. The most common strategy for identifying possible regulatory elements has hitherto been inter-species conservation studies. Using this approach, all elements flanking the gene that are conserved more than some threshold are retained for further experimental characterization. The reason underlying this strategy is that truly functional elements are under evolutionary pressure to retain their function across species. Given the technical complexity of associated transgenic experiments, this turns out to be a fairly inefficient strategy, especially since the number of candidate regulatory elements increases as larger genomic regions are examined (to account for distal regulation). 
Such a scenario prompts the need for an integrative strategy to reduce the number of candidates obtained from a purely conservation-based search strategy using other, complementary genomic modalities.

Recently, \cite{Khandekar2004} reported the characterization of two enhancer elements, conferring urogenital-specific expression of \textit{Gata2}, between $80-150$kb away from the gene locus, on chromosome $6$. In this work, we examine if genomic features, other than sequence identity, are predictive of the location of these elements. These feature span sequence, expression and interactome perspectives for such regulatory elements. We will also attempt to motivate the utility of these approaches (metrics and data sources) as well as their biological relevance alongside (how they fit into the biophysics of transcriptional regulation). It must be pointed out that there is large paucity in data availability, in that data specific to the developing kidney is hard to come by. Under this constraint, we have made some biologically plausible assumptions so as to obtain maximum information from currently available data sources.

\section{Rationale and Data Sources:} \label{data_sources}

The overall schematic of distal transcriptional regulation via looping is given in Fig. \ref{fig:transcription2}. This schematic suggests the decomposition of the regulatory process along three main modalities: sequence, expression and interactome. Our main goal in this paper is to understand urogenital (kidney) enhancer behavior from these three perspectives. These attributes are discussed below:

\begin{figure}[!h!t!b]
\caption{Distal enhancer-promoter interaction via looping is mediated via protein interactions during TF complex formation. The set of TFs that are putatively recruited at the proximal promoter and distal enhancer can be found from sequence and expression data \cite{CSB2007}. Evidence of protein-interaction between proximal and distal TFs can be found from interaction databases.}\label{fig:transcription2}
\end{figure}

\begin{enumerate}
\item
\textbf{Sequence Perspective:}
To build motif signatures underlying kidney-specific enhancer activity, it would be best to have a database of previously characterized urogenital (UG) enhancers. However, due to the unavailability of such data, we utilize kidney-specific promoter sequences and histone-modified sequences of enhancers to find motif-signatures of regulatory elements that are potentially UG enhancers. 
\begin{itemize}
\item
\underline{Promoters of kidney-specific genes}:
A catalog of kidney-specific mouse promoters is available from the GNF Symatlas (\emph{http://symatlas.gnf.org/}). This database contains list of annotated genes and their expression in several tissue types, including the kidney. Since the proximal promoter of such kidney-specific genes harbors the transcriptional machinery for gene regulation, their sequences putatively have motifs that are associated with kidney-specific expression. Additionally, promoters that are spatio-temporally expressed during kidney development are also analyzed (MGI: \emph{http://www.informatics.jax.org/}). The GNF dataset profiles mostly adult tissue-types. Since our goal is to study enhancer activity during nephrogenesis, we focus on genes expressed between day $e10$ and $e12$ in the developing kidney - such a list is obtained from the MGI database.

Without loss of generality, we use six-nucleotide motifs (hexamers) as the motifs. This is based on the observation that most transcription factor binding motifs have a $5-6$ nucleotide core sequence with degeneracy at the ends of the motif. A similar setup has been introduced in (\cite{DrosophilaFeatureDiff}, \cite{PromFind}). The main difference in our approach from such previous work is that differential hexamer analysis was done for the same class of sequences, and the statistical nature of the ``test-set" is, by design similar to the training set. That is, in \cite{DrosophilaFeatureDiff}, differential hexamers are found between known Cis-Regulatory Modules (CRMs) and non-CRMs, and used for the prediction of new CRMs from sequence. On the other hand, \cite{PromFind} deals with finding hexamer features of known promoters and using them to predict new promoters from sequence. In our case, however, we don't have enhancer data (equivalent to CRMs) and we are using promoter-data for the prediction of enhancer (CRM) instead. Thus, the nature of the test sequence is very different. We demonstrate that our approach is partially useful in the discovery of putative enhancers from sequence.
Also,the presented motif-finding approach does not depend on motif length and can be scaled depending on biological knowledge.

We set up the motif discovery as a feature extraction problem from these tissue-specific promoter sequences and then build  a random forest (RF) classifier to classify new sequences into specific and non-specific categories based on these identified sequence features (motifs). Based on the motifs derived using a RF classifier algorithm we are able to accurately classify more than $95\%$ (training-error rate) of tissue-specific genes based upon their upstream promoter region sequences alone. Since promoters are non-coding regulatory regions, the derived motifs can be putatively used to find kidney-specificity of other non-coding regions genome-wide (Section: \ref{kidney-RF}).
\item
\underline{Chromatin marks in known regulatory elements}: Using the recently released ENCODE data, a catalog of sequences that undergo histone modifications such as methylation and acetylation is available for analysis \cite{Sanger-Histone}. Reports suggest that mono-methylation of the lysine residue of Histone $H3$ is associated with  enhancer activity \cite{chromatin-ChIP} whereas tri-methylation of $H3K4$ and $H3$ acetylation are associated with promoter activity. Using this set of $H3K4me1$, $H3K4me3$ and $H3ac$ sequences, we aim to find sequence motifs that are indicative of such epigenetic preferences during transcription. Though epigenetic data is available for five different cell lines, we choose the HeLa cell line data because of its widespread use as a model system to understand transcriptional regulation \textit{in-vitro} in the laboratory.
Thus, we build a RF classifier to discriminate monomethylated $H3K4$ sequences from trimethylated $H3K4$/acetylated $H3$ sequences. We note that this data is \emph{not} kidney-specific, and such data is yet to become available. This yields motifs associated with epigenetic properties of promoters and enhancers, which are potentially predictive of the regulatory potential for novel sequences (section: \ref{histone-RF}).
\end{itemize}
\item
\textbf{Expression Perspective:}
There is limited expression data for the developing mouse kidney, mainly due to technical reasons concerning small tissue yield at such early time points. For this study, we use microarray expression data from a public repository of kidney microarray data (\emph{http://genet.chmcc.org} \cite{Stuart2003}, \emph{http://spring.imb.uq.edu.au/} \cite{GrimmondLittle2005}). Each of these sources contain expression data profiling kidney development from about day $10.5$ dpc to the neonate stage. Such expression data can be mined for potential regulatory influence between upstream TF genes and \textit{Gata2}.
\begin{itemize}
\item
\textit{Inference of TF effectors at the promoter region}:
The TFs putatively recruited at the proximal promoter are identified using the Directed Information metric, that uses gene-expression level influence in addition to phylogenetic conservation of the corresponding binding site. We have earlier shown that DTI is a good predictor of gene influence and can be used to infer transcriptional regulatory networks \cite{CSB2007}. A more detailed explanation is given in sections: \ref{prom_TF} through \ref{boot_CI}.
\item
\textit{Inference of TF effectors at each non-coding region}:\\
At the distal enhancer, it is believed that there is recruitment of tissue-specific transcription factors that co-operate with the basal transcriptional machinery (at the promoter) to direct tissue-specific gene expression (\cite{Kleinjan2005}, \cite{Fraenkel2006}). Whereas phylogeny and expression-based influence metrics can yield high confidence candidates for promoter TFs, 
a similar analysis for enhancers is not possible, because of higher order effects (\cite{Meyer2004}, \cite{Kreiman2004}). To this end, the only plausible way to search for enhancer TFs is to combine phylogeny with tissue-specific annotation (from UNIPROT or MGI). Hence, every transcription factor, whose motif is conserved at a non-coding (putative enhancer) region and is tissue-specific in annotation is considered a likely candidate TF at that non-coding region.
\end{itemize}
\item
\textbf{Interactome Perspective:}
The identification of phylogenetically conserved effector TFs at the promoter (identified via DTI), as also those that are phylogenetically conserved at the putative enhancers, lead to the exploration of protein-interactions between these TFs, during distal enhancer-promoter interaction (Sec:\ref{ppi_meta}). The STRING database (\emph{http://string.embl.de}) integrates various experimental modalities (genomic context, high-throughput experiments such as co-immunoprecipitation, co-expression and literature) to maintain a list of organism-specific functional protein-association networks that is amenable to such exploration. 
\end{enumerate}

In this work, the above questions will be integratively answered for training data as well as in the context of the urogenital enhancers identified in \cite{Khandekar2004}. We aim to show that each of these `features' have a predictive value for the identification of enhancers and the integration of these heterogeneous data can lead to potential reduction in false positive rate during large-scale enhancer discovery, genome-wide. To date, there has been no comprehensive study for summarizing these various heterogeneous data sources to understand transcriptional regulation.

\section{Validation/Biological Application}

As suggested in Sec: \ref{introduction}, we use the recently identified \textit{Gata2} urogenital (UG) enhancers to validate our computational approach. All the data sources (and their analysis) are therefore going to be focused on the kidney.

The experimental characterization of these enhancers was done as follows. Based on BAC transgenic \cite{Khandekar2004} studies, the approximate location of the urogenital enhancer(s) of \textit{Gata2} were localized to a $70$ kilobase region on chromosome $6$. Using inter-species conservation plots, four elements were selected for transgenic analysis in the mouse. These were designated UG$1$, $2$, $3$ and $4$. After a lengthy and resource-intensive experimental effort, two out of these four non-coding elements, $UG2$ and $UG4$ were found to be true UG enhancers. Our goal is to find ``features" at the sequence, expression and interactome level, that are predictive of this reported behavior of elements $UG1-4$ in the developing kidney.

It is easy to see the utility of such a methodology, since this can be scaled up contextually for other genes of interest. Given the complexity of $1\%$ of the genome, made possible by the ENCODE project, the search for functional elements genome-wide is going to be an important and challenging exercise.

\section{Organization}

With a view to understanding the elements of transcriptional regulation, the first part of this paper (Sections \ref{data_processing}-\ref{histone-RF}) addresses the problem of identifying motif signatures representative of transcriptional control from kidney-specific promoters and epigenetically marked sequences. The second part of this work (Sections \ref{prom_TF} - \ref{enh_TF}) integrates phylogeny and expression data to find regulatory TFs at the proximal promoter and enhancer(s) of \textit{Gata2}. Using the notion of TF interactions between enhancer and promoter, we examine if protein-interaction data (Sec: \ref{ppi_part2}) can offer supporting evidence for the observed \textit{in-vivo} behavior of four putative \textit{Gata2} regulatory elements. Classifiers are designed to discriminate regulatory vs. non-regulatory regions based on each of these multiple modalities. Finally, a probabilistic combination of these classifiers is done to obtain a validation (Sec: \ref{data_integration}) of the \textit{Gata2} UGEs ($UG1-4$). Sections: \ref{summary} and \ref{conclusions} conclude the paper.

\section{Sequence Data Extraction and Pre-processing}\label{data_processing}

The Novartis foundation tissue-specificity atlas [\emph{http://symatlas.gnf.org/}], has a compendium of genes and their corresponding tissues of expression. Genes have been profiled for expression in about twenty-five tissues, including adrenal gland, brain, dorsal root ganglion, spinal chord, testis, pancreas, liver etc. Considering these diversity of tissue-types, one concern with the interpretation of this data is the variability in expression across tissue-types. To address this concern, we take a fairly stringent approach - if a gene is expressed in less than three tissue types, it is annotated tissue-specific (\textit{`ts'}), and if it is expressed in more than $22$ tissue types, it is annotated to be non-specific (\textit{`nts'}). Based on this assignment, we find a list of $86$ genes that are tissue-specific as well as have kidney expression (MGI: \emph{http://www.informatics.jax.org/}). For these kidney-specific genes, we extract their promoter sequences from the ENSEMBL database (\emph{http://www.ensembl.org/}), using  sequence $2000$bp upstream and $1000$bp downstream up to the first exon relative to the transcriptional start site reported in ENSEMBL (release 37).

Before proceeding to motif selection, a matrix of motif-promoter correspondences is created. In this matrix, the counts of hexamer (six-nucleotide) motif occurrence in the \textit{`ts'} and
\textit{`nts'} promoters is obtained using sequence parsing (\textit{R package: `seqinr'}). The motif length of six is not overly restrictive, since it corresponds to the consensus binding site size of several
annotated transcription factor motifs in the TRANSFAC/JASPAR databases. A Welch t-test is then performed between the relative counts of each hexamer in the two expression categories (\textit{`ts'} and \textit{`nts'}) and the top $1000$ hexamers with $p-value \le 10^{-6}$ are selected. This set of discriminating hexamers is designated ($\overrightarrow{\textbf{H}} =
H_1,H_2,\ldots,H_{1000}$). This procedure resulted in two hexamer-gene co-occurrence matrices, - one for the \textit{`ts'} (or $+1$) class of dimension $N_{train,+1} \times 1000$ and the
other for the \textit{`nts'} (or $-1$) class - dimension $N_{train,-1} \times 1000$. Here $N_{train,+1}$ is the matrix of the $86$ kidney-specific genes. $N_{train,-1}$ is the set of \textit{`nts'} that do not have kidney-specific expression.

As an illustration, we show a representative matrix (Table. \ref{motif_promoter_mtx}).
\begin{table}[!h!b!t]
\centering
\begin{tabular}{cccccc}
Ensembl Gene ID & AAAAAA    & AAATAG  & Class \\
ENSG00000155366 & 1      & 1   &  +1 \\
ENSG000001780892 & 4   & 3 &   +1 \\
ENSG00000189171 & 1  &  2 &    -1 \\
ENSG00000168664 & 4  &  3 &   -1 \\
ENSG00000160917 & 2  &  1 &    -1 \\
ENSG00000176749 & 1  &  1 &   -1 \\
ENSG00000006451 & 3  &  2 &   +1
\end{tabular}
\caption{The `motif count matrix' for a set of gene-promoters.
The first column is their ENSEMBL gene identifiers, the next $2$
columns are hexamer quantile labels, and the last column is the
corresponding gene's class label ($+1/-1$).}\label{motif_promoter_mtx}
\end{table}

All the above steps, from sequence extraction, parsing and quantization to obtain hexamer-promoter counts that are done for the kidney-specific genes can be repeated for the histone-modified sequences. This dataset is obtained from the Sanger ENCODE database (\emph{http://www.sanger.ac.uk/PostGenomics/encode/data-access.shtml}), and contains $298$ sequences that undergo modification ($m1/me3/ac$) in histone ChIP assays. $140$ of these correspond to $H3K4me1$ (enhancers), and $158$ correspond to $H3K4me3/H3ac$ marks (promoters). Here, the $1000$ hexamers discriminating $H3K4me1$-sequences ($+1$ set) and a $(H3K4me3/H3ac)$ ($-1$), are designated $\overrightarrow{\textbf{H'}}=H'_1,H'_2,\ldots,H'_{1000}$.

\begin{table}[!h!b!t]
\centering
\begin{tabular}{cccccc}
Sequence & AAAATA & AAACTG & Class\\
chr2:41410492-41411867 & 2 & 1 & +1 \\
chr6:41654502-41654782 & 4 & 2 & +1 \\
chr3:41406971-41408059 & 1 & 1 & -1 \\
chr2:41665970-41667002 & 2 & 3 & +1 \\
chr4:41476956-41478365 & 1 & 2 & -1 \\
chr5:41530471-41531046 & 2 & 2 & -1 \\
chrX:41783327-41784532 & 1 & 2 & +1
\end{tabular}
\caption{The `motif count matrix' for a set of histone-modified sequences.
The first column is their genomic locations along the chromosome, the next $2$
columns are hexamer quantile labels, and the last column is the
corresponding sequence class label ($+1/-1$).}\label{motif_histone_mtx}
\end{table}

\section{Motif-Class Correspondence Matrices}

From the above, $N_{train,+1} \times 1000$ and $N_{train,-1} \times 1000$ dimensional count matrices are available both for the kidney-promoter and histone-modified sequences. Before proceeding to the feature (hexamer motif) selection step, the counts of the $M = 1000$ hexamers in each training sample need to be normalized to account for variable sequence lengths.
In the co-occurrence matrix, let $gc_{i,k}$ represent the absolute count of the $k^{th}$ hexamer, $k \in {1,2,\ldots,M}$ in the $i^{th}$ gene. Then, for each gene $g_{i}$, the quantile labeled matrix has $X_{i,k} = l$ if $gc_{i,[\frac{l-1}{K}M]} \le gc_{i,k} < gc_{i,[\frac{l}{K}M]}, K=4$. Matrices of dimension $N_{train,+1} \times 1001$, $N_{train,-1} \times 1001$ for the specific and non-specific training samples are now obtained. Each matrix contains the quantile label assignments for the $1000$ hexamers $(X_i, i \in (1,2,\ldots,1000))$, as stated above, and the last column would have the corresponding class label ($Y = -1/+1$).
Having constructed two groups of genes for analysis, tissue specific (\textit{`ts'}) and non-tissue specific (\textit{`nts'}) - we seek to find hexamer motifs which are most discriminatory between these two classes. Our goal would be to make this set of motifs as small as possible - i.e. to achieve maximal class partitioning with the smallest feature subset. Towards this goal, we explore the use of random forests (RF) \cite{Breiman_RF} for finding such a discriminative hexamer subset.

\section{Random Forest Classifiers}

A random forest (RF) is an ensemble of classifiers obtained by aggregating (bagging) several classification trees  (\cite{HastieTibshirani2002}, \cite{Breiman_RF}). Each data point (represented as an input vector) is classified based on the majority vote gained by that vector across all the trees of the forest. Each tree of the forest is grown in the following way:
\begin{itemize}
\item
A bootstrapped sample (with replacement) of the training data is used to grow each tree. The sampling for bootstrapped data selection is done individually at each tree of the forest.
\item
For an $M$-dimensional input vector, a random subspace of $m$ ($\ll M$)-dimensions is selected, and the best split on this subspace is used to split the node. This is done for all nodes of the tree. Each tree is grown to maximum length, with no pruning.
\end{itemize}

During the training step, before sampling by replacement, one-third of the cases is kept ``out of the training bag". This oob (out-of-bag) data is used to obtain an unbiased estimate of the classification error as trees are added to the forest. It is also used to get estimates of variable importance.

From the above we see that the classifier structure of the random forest is an ensemble of trees. Each tree is trained and built on a different bootstrap sample (split) of the training data. Hence each tree has a different topology. Unlike a tree classifier, therefore, it is not possible to obtain a ``consensus topology" of the RF classifier. In the absence of one unifying structure for the purpose of visualization, we can inspect the other outputs like variable importance, confusion matrix, and OOB error rate to ascertain the accuracy and performance of the RF classifier.

The variables selected for optimal partitioning over class labels can be examined from a variable importance plot which indicates which variables are most discriminatory between these two classes (\cite{Breiman_RF}, \cite{RandomForest_doc}). It is also to be noted that random forests afford the dual advantage of both training and test-set error estimation (through the OOB data) during the overall training procedure. Thus there is no separate procedure for test-set error estimation that needs to be implemented in the case of RFs. Each tree in the ensemble is trained on a $\frac{2}{3}rd-\frac{1}{3}rd$ split of the data. Each tree is grown to get the least oob error before being incorporated into the classifier ensemble. 


A confusion matrix is one representative tool to understand the performance of the RF classifier. After the training process, the confusion matrix measures the discordance between true and predicted classes (and can be used for OOB error estimation). Each row represents the instances of the actual class, while each column of the matrix represents the instances in a predicted class. The matrix can then be used for false-positive, false-negative, true-negative and true-positive rate computations.




Several interesting insights into the data are available using random forest analysis. The variable importance plot yields the variables that are most discriminatory for classification under the `ensemble of trees' classifier. This importance is based on two measures- `Gini index' and `decrease in accuracy'. The Gini index is an entropy based criterion which measures the purity of a node in the tree, while the other metric simply looks at the relative contribution of each variable to the accuracy of the classifier. 
For our studies, we use the `randomForest' package for R \cite{RandomForest_doc}. The classifier performance on the individual data and the related diagnostics are mentioned under each head (Secs: \ref{kidney-RF} and \ref{histone-RF}).

\section{Random Forests on Kidney-specific promoters}\label{kidney-RF}
In this section, we aim to find discriminating sequence motifs between a set of kidney-specific promoters and housekeeping promoters with a goal to find sequence motifs underlying kidney-specific regulation. The kidney enriched dataset has $86$ genes that are assigned to a tissue specific class and have higher than mean expression in the kidney. For the purpose of training and testing, we consider the set of housekeeping genes identified from the \textit{`nts'} class and reported in literature (\cite{hkg1}, \cite{hkg2}). 
There are almost $1500$ genes in the housekeeping gene (\textit{`nts'}) set. Since, this would lead to unbalanced predictions during classifier training, we use a stratified sampling approach \cite{RandomForest_doc} to select for a sample size that reduces this effect (the sampling itself is done with a prior on the relative sizes of the two classes). Here, the set of $(-1)$ promoter-sequences are taken to be of the same size as the $(+1)$ class. Using this approach, we obtain a training-error classification accuracy of $> 95\%$ on the kidney enriched tissue-specificity data set.

\begin{figure}[!h!t!b]
\centerline{\includegraphics[width=3.2in,height=3in]{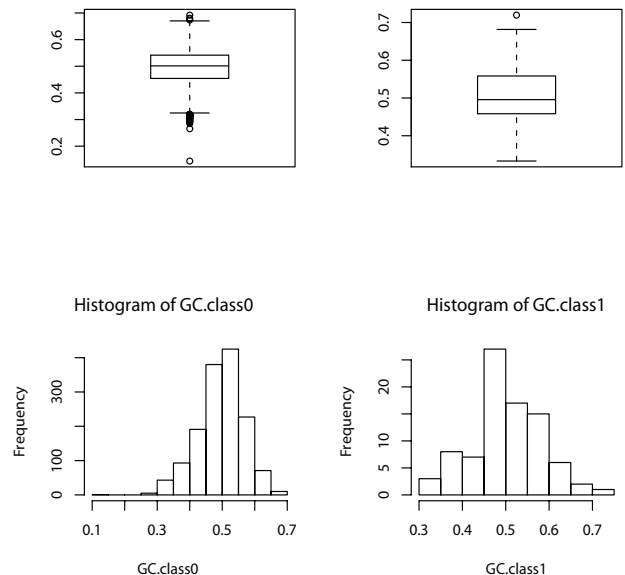}}
\caption{GC plots for sequence bias in kidney-specific vs. housekeeping promoters.}\label{fig:kidney_gc}
\end{figure}

\begin{figure}[!h!b!t]
\centerline{\includegraphics[width=3.2in,height=3in]{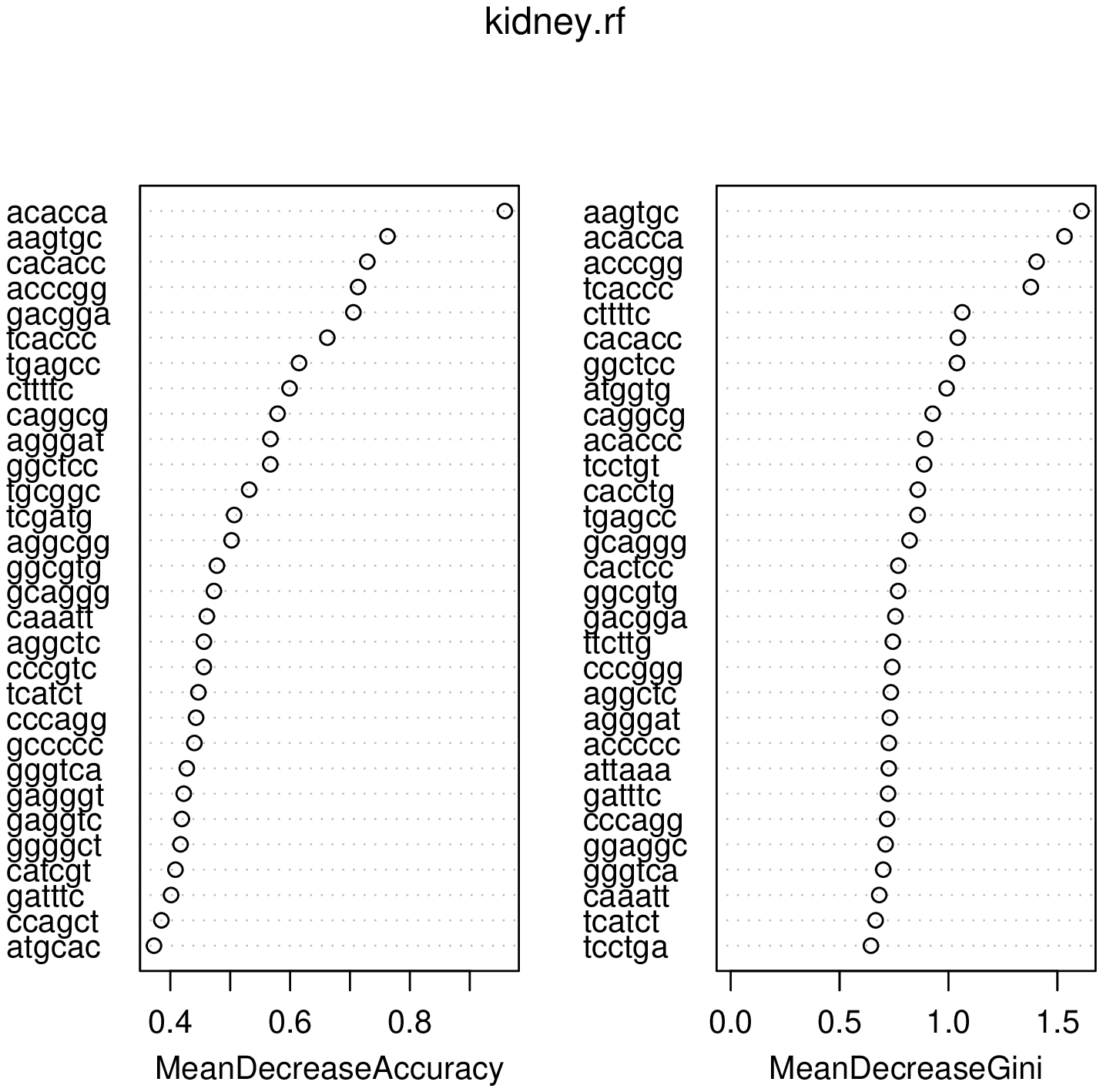}}
\caption{Top hexamers which can discriminate between kidney-specific
and house-keeping genes.}\label{fig:kidney_hexamers}
\end{figure}

Before proceeding to motif identification, it is necessary to check for possible sequence bias (GC composition) between the two classes of promoters (kidney-specific vs. housekeeping). Though there are several kinds of sequence bias \cite{seq_bias}, the composition bias is most closely related to this problem. If there is a significant bias, then the motifs turn out to be just GC rich sequences that are not very biologically informative \cite{DWE2005} for regulatory potential. The GC composition of these two classes of sequences is represented in Fig. \ref{fig:kidney_gc}.  We note that though only a subset of `nts' gene-promoters were used during the RF analysis, we show the GC-composition for the entire class of `nts' sequences for completeness. As can be seen, the average GC composition is the same. The ROC space representation and variable importance plot for the overall classification is indicated below (Fig. \ref{fig:roc_rf} and Fig. \ref{fig:kidney_hexamers}). The confusion matrices are all explained in the context of the classifier combination in Section:\ref{data_integration}.

To address a related question, we examine if the top ranked hexamers in the kidney dataset correspond sequence-wise to known transcription factor binding sites. Using the publicly available Opossum tool (\textit{http://www.cisreg.ca/cgi-bin/oPOSSUM/opossum/}) or MAPPER (\emph{http://bio.chip.org/mapper}), we found several interesting transcription factors to map to these motifs, such as \textit{Nkx}, \textit{ARNT}, \textit{c-ETS}, \textit{FREAC4}, \textit{NFAT}, \textit{CREBP}, \textit{E2F}, \textit{HNF4A}, \textit{Pax2}, \textit{MSX1}, \textit{SP1} several of which are kidney-specific. Though this is highly consistent with the tissue-specificity of the dataset, the functional relevance of  these sites remains to be experimentally validated.

\section{RFs on chromatin-modified sequences}\label{histone-RF}

We train a RF classifier on a set of $298$ sequences from chromosome sequence that have varying histone modifications associated with them (namely, $H3K4me1/me3$, and $H3ac$ ), as mentioned in Section: \ref{data_sources}. These sequences had a high level of the corresponding histone-modification from ChIP experiments. The other regions that were assayed for but did not have high levels of modification are not considered in this analysis. These are derived from the HeLa cell line and are not necessarily context-specific for kidney development. However, given the widespread  use of this cell line for transcriptional studies, we aim to find if the motifs associated with regulatory elements are indeed predictive of enhancer activity.

\begin{figure}[!t!h!b]
\centerline{\includegraphics[width=3.2in,height=3in]{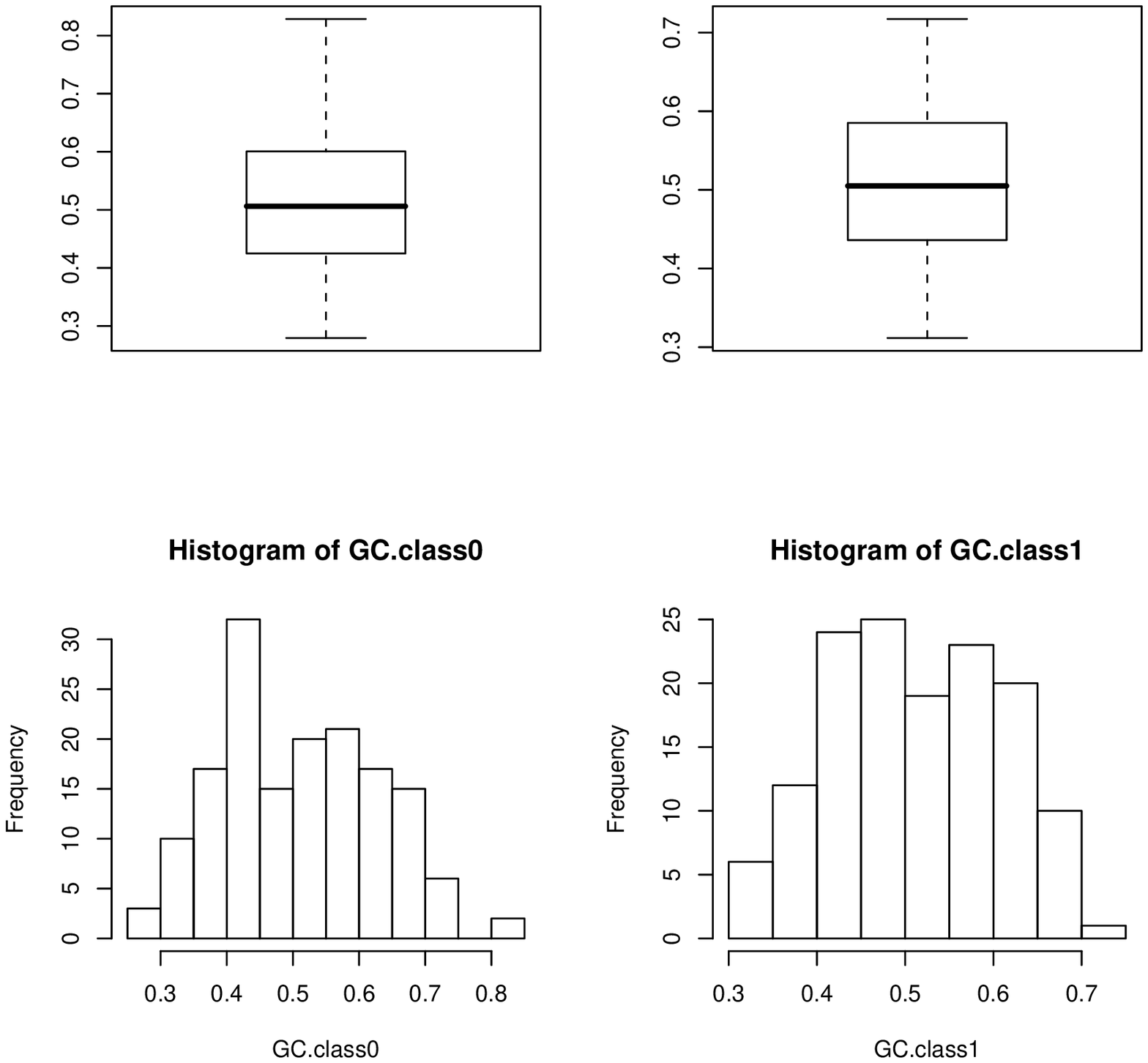}}
\caption{GC plots for sequence bias in $H3K4me1$ histone sequences vs. $H3K4me3$ and $H3ac$  sequences.}\label{fig:histone_gc}
\end{figure}

Here too, we examine the GC-composition bias of these two sequence classes (Fig. \ref{fig:histone_gc}) and confirm that there is no such sequence bias that would skew the discovery and subsequent interpretation of these epigenetic motifs.

\begin{figure}[!t!h!b]
\centerline{\includegraphics[width=3.2in,height=3in]{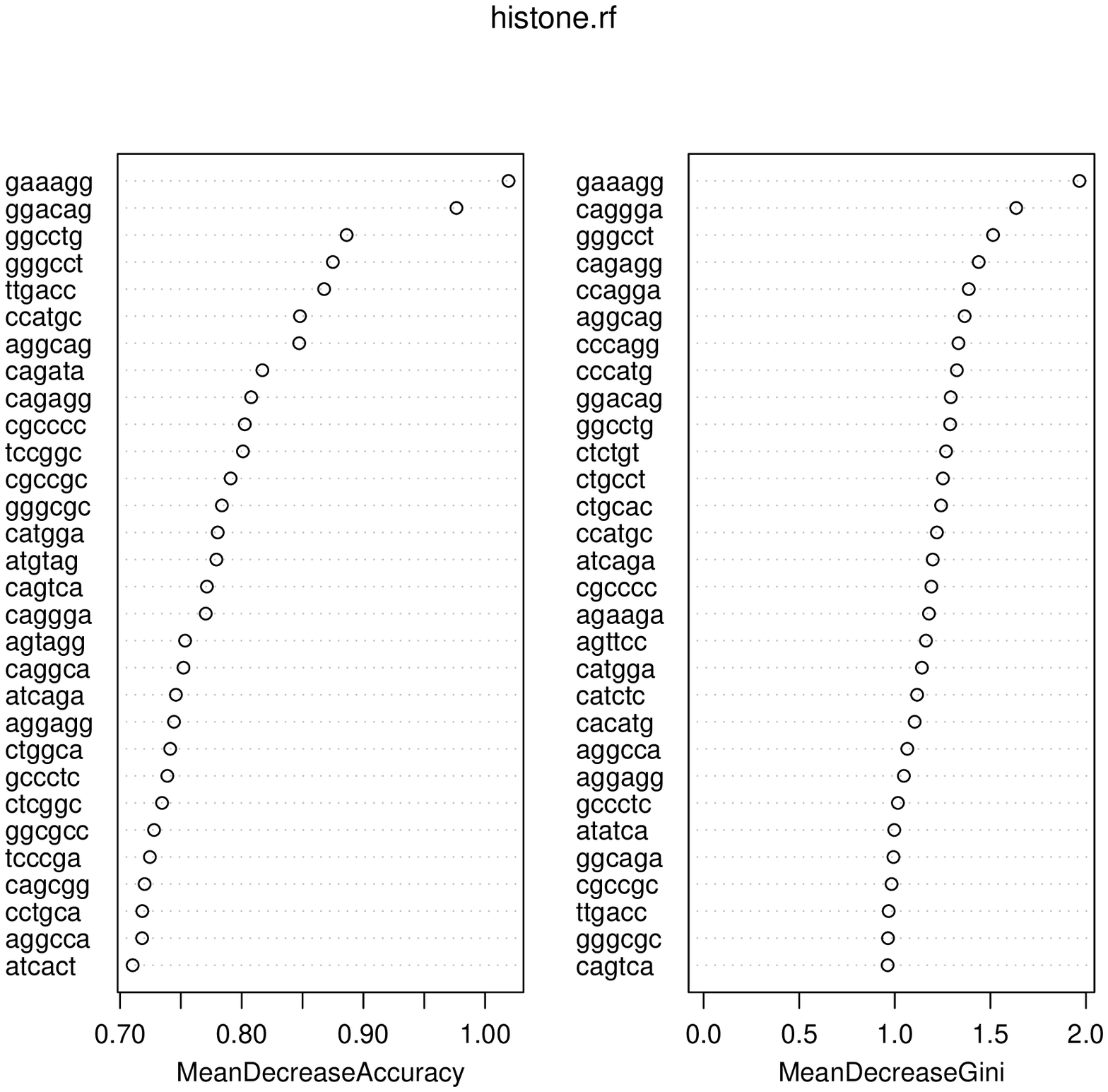}}
\caption{Top hexamers which can discriminate between $H3K4me1$ histone sequences vs. $H3K4me3$ and $H3ac$ sequences.}\label{fig:histone_hexamers}
\end{figure}

The motifs obtained from the random forest analysis indicate the ``sequence-preferences" of regulatory elements that are kidney-specific (Fig. \ref{fig:kidney_hexamers}) or nucleosome-free (Fig. \ref{fig:histone_hexamers}). For the kidney-specific case, the underlying caveat is that co-expression does not imply co-regulation; however we are only using the discovered motifs to understand the ``sequence-preferences" of kidney-specific regulatory-regions \cite{RPscore} rather than using them for \textit{de-novo} prediction of new genes that are regulated by the same transcriptional machinery. Most of the motifs do not overlap TFBS motifs and might be indicative of more interesting sequence properties.
We analyze the performance of these classifiers on the $4$ UG enhancers, mentioned previously. In both cases, $UG2-4$ are classified as kidney-specific enhancers, whereas $UG1$ is correctly classified as not being regulatory. Additionally, a control set of ``promoter-independent" enhancers derived from the Mouse Enhancer database \cite{EnhancerBrowser} was also classified as enhancers based on these chromatin signatures. This high prediction accuracy inspite of non-specificity of cell context (\textit{HeLa} cell line)
is very interesting and has potentially high predictive value. This is explored further in Sec: \ref{data_integration}.

We now proceed to the mechanistic insight (based on TF effector identification and PPI) mentioned in Section. \ref{introduction} to understand the behavior of putative regulatory elements.

\section{PPI between promoter and enhancer TFs}\label{ppi_meta}

In order to understand the nature of distal interactions between the enhancer and promoter TFs (Fig. \ref{fig:transcription2}), we decouple the overall regulation problem into three parts:
\begin{enumerate}
\item Identification of putative TF effectors at the promoter (Section: \ref{prom_TF}),
\item Identification of enhancer TFs (Section: \ref{enh_TF}), and
\item Examination of the interaction-graph formed between enhancer-TFs and promoter TFs (Section: \ref{ppi_part2}).
\end{enumerate}

\subsection{TF effector identification at Promoter and Enhancer}\label{prom_TF}
\emph{Promoter TF identification:}
TFs that regulate basal transcription at the promoter can be identified from phylogenetic conservation or co-expression studies. In this approach, the promoter sequence (here, the \textit{Gata2} promoter) is aligned across multiple species and the TFBS motifs that are conserved in the multiple alignment are considered to be putative effectors of gene regulation. An additional step involves examining the promoters of all genes that are co-expressed in the same spatio-temporal manner as the gene of interest (e.g.: \textit{Gata2} in the kidney). Such sequence-based approaches have been examined in literature (\cite{Fraenkel2006}, \cite{Kreiman2004}, \cite{Meyer2004}).

Since the list of putative TFs (identified above) that potentially bind at the promoter is still large, there have been efforts to incorporate gene-expression data to reduce the set of potential TF effectors. In this scenario, if the gene corresponding to the conserved TF has a high expression-level influence on \textit{Gata2} expression, then that TF has stronger evidence for being a potential regulator (\cite{ARACNE}, \cite{CoD}).
Recently, we introduced the directed information (DTI) as a metric to infer expression-level influence between any putative transcription factor (TF) gene and a target gene (such as \textit{Gata2}) \cite{CSB2007}. We will briefly summarize the utility of DTI for TF effector identification in the following sections (Sec. \ref{dti_formulation} and \ref{boot_CI}). This seeks to integrate sequence and expression data into the determination of relationships between transcription factors and their target-genes. All additional details (performance on synthetic data, other biological data and comparison with other metrics) are available in \cite{CSB2007}. Information-based measures have enabled the investigation of non-linear gene relationships in the presence of measurement noise \cite{ARACNE}. An important point to note is that unlike mutual information, the DTI is a \textit{directed} metric that enables the inference of both strength and direction of gene influence.

\begin{figure}[H]
\centerline{\includegraphics[width=3.2in,height=2.2in]{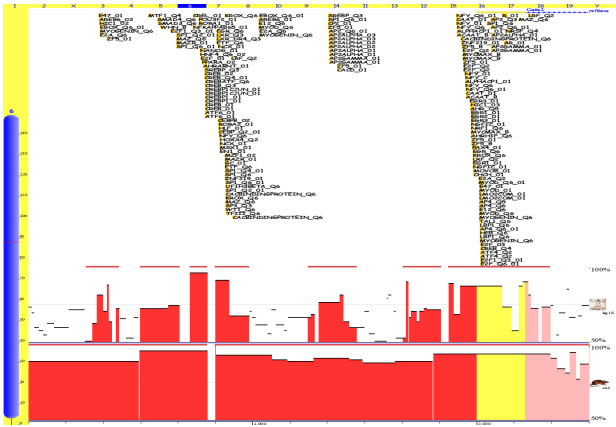}}
\caption{TFBS conservation between Human, Mouse and Rat, upstream
(x-axis) of \textit{Gata2}, from
\emph{http://www.ecrbrowser.dcode.org/}. The mouse sequence is the base sequence and is hence not displayed. The dark and light red regions correspond to potential TF binding regions on DNA.}\label{fig:TFBS_conservation}
\end{figure}

\subsubsection{DTI Formulation}\label{dti_formulation}
As alluded to above, there is a need for a viable influence metric that can find relationships between the TF ``effector" gene (identified from phylogenetic conservation) and the target gene (like \textit{Gata2}). Several such metrics have been proposed, notably correlation, coefficient of determination (CoD), mutual information etc. To alleviate the challenge of detecting non-linear gene interactions, an information theoretic measure
like mutual information has been used to infer the conditional dependence among genes by exploring the structure of the joint distribution of the gene expression profiles
\cite{ARACNE}. However, the absence of a directed dependence metric has hindered the utilization of the full potential of information theory. In this section, we examine the
applicability of one such metric - the directed information criterion (DTI), for the inference of non-linear, directed gene influences. 

The DTI - which is a measure of the directed dependence between two $N$-length random processes $X \equiv X^N$ and $Y \equiv Y^N$ is given by \cite{Massey1990}:
\begin{gather*}
I(X^N \rightarrow Y^N) = \sum_{n=1}^{N}I(X^n;Y_n|Y^{n-1})\tag{$1$}
\label{eq1}
\end{gather*}

Here, $Y^n$ denotes $(Y_1, Y_2,.., Y_n)$, i.e. a segment of the realization of a random process $Y$ and $I(X^N;Y^N)$ is the Shannon mutual information \cite{Cover&Thomas}.

An interpretation of the above formulation for DTI is in order. To infer the notion of influence between two time series (mRNA expression data) we find the mutual information between the entire evolution of gene $X$ (up to the current instant $n$) and the current instant of $Y$ ($Y_n$), given the evolution of gene $Y$ up to the previous instant $n-1$ (i.e. $Y^{n-1}$). This is done for every instant, $n \in (1,2,\ldots,N)$, in the $N$ - length expression time series. 

As already known, $I(X^N; Y^N) = H(X^N)-H(X^N|Y^N)$, with $H(X^N)$ and $H(X^N|Y^N)$ being the entropy of $X^N$ and the conditional entropy of $X^N$ given $Y^N$, respectively. Using this
definition of mutual information, the DTI can be expressed in terms of individual and joint entropies of $X^N$ and $Y^N$. The task of $N$-dimensional entropy estimation is an important one and due to computational complexity and moderate sample size, histogram estimation of this multivariate density is unviable. However, several methods exist for consistent entropy estimation of multivariate small sample data (\cite{ErikLearnedMiller2003}, \cite{NemenmanBialek2002}, \cite{Paninski2003}, \cite{WilettNowak2004}). In the context of microarray expression data, wherein probe-level and technical/biological replicates
are available, we use the method of \cite{ErikLearnedMiller2003} for entropy estimation.

From $(1)$, we have,\\
\begin{gather*}
I(X^N \rightarrow Y^N) = \sum_{n=1}^N [H(X^n|Y^{n-1})-H(X^n|Y^{n})] \\
= \sum_{n=1}^N \{ [H(X^n,Y^{n-1})-H(Y^{n-1})]-\\
[H(X^n,Y^{n})-H(Y^{n})]\} \tag{$2$}
\end{gather*}

\begin{itemize}
\item
To evaluate the DTI expression in $Eqn. 2$, we need to estimate the entropy terms $H(X^n,Y^{n-1})$, $H(Y^{n-1})$, $H(X^n,Y^{n})$ and $H(Y^{n})$. This involves the estimation of marginal and joint entropies of $n$ random variables, each of which are $R$ dimensional, $R$ being the total number of replicates (probe-level, biological and technical).
\item
Though some approaches need the estimation of probability density of the $R$-dimensional multivariate data $(X^n)$ prior to entropy estimation, one way to circumvent this is to the use the method proposed in \cite{ErikLearnedMiller2003}. This approach uses a Voronoi tessellation of the $R$-dimensional space to build nearly uniform partitions (of equal mass) of the density. The set of Voronoi regions $(V^1,V^2,\ldots, V^n)$ for each of the $n$ points in $R$-dimensional space is formed by associating with each point $X_k$, a set of points $V^k$ that are closer to $X_k$ than any other point $X_l$, where  the subscripts $k$ and $l$ pertain to the $k^{th}$ and $l^{th}$ time instants of gene expression. 
\item
Thus, the entropy estimator is expressed as : $\hat{H}(X^n) =
\frac{1}{n} \sum_{i=1}^{n}\textrm{log}(nA(V^{i}))$, where $A(V^{i})$ is the $R$-dimensional volume of Voronoi region $V^{i}$. $A(V^i)$ is computed as the area of the polygon formed by the vertices of the convex hull of the Voronoi region $V^i$. This estimate has low variance and is asymptotically efficient \cite{ErikLearnedMiller2004}.
\end{itemize}

To obtain the DTI between any two genes of interest ($X$  and $Y$) with $N$-length expression profiles $X^N$ and $Y^N$ respectively, we plug in the entropy estimates computed above into the expression ($2$).

From the definition of DTI, we know that $0 \le I(X_i^N \rightarrow Y^N) \le I(X_i^N;Y^N) < \infty$. For easy comparison with other metrics, we use a normalized DTI metric given by $\rho_{DI}= \sqrt{1-e^{-2 I(X^N \rightarrow Y^N)}} = \sqrt{1-e^{-2\sum_{i=1}^N I(X^i;Y_i|Y^{i-1})}}$. This maps the large range of DI, ($[0,\infty]$) to lie in $[0,1]$. Another point of consideration is to estimate the significance of the `true' DTI value compared to a null distribution on the DTI value (i.e. what is the chance of finding the DTI value by chance from the series $X$ and $Y$). This is done using empirical $p$-value estimation after bootstrap resampling (Sec: \ref{boot_CI}). A threshold $p$-value of $0.05$ is used to estimate the significance of the true DTI value in conjunction with the density of a random data permutation, as outlined below.

\subsubsection{Significance Estimation of DTI}\label{boot_CI}
We now outline a procedure to estimate the empirical $p$-value to ascertain the significance of the normalized directed information $\hat{I}(X^N \rightarrow Y^N)$ between any two $N$-length time series $X \equiv X^N=(X_1,X_2,\ldots,X_N)$, and $Y \equiv Y^N=(Y_1,Y_2,\ldots,Y_N)$. In our case, the detection statistic is $\Theta = \hat{I}(X^N \rightarrow Y^N)$ and the chosen acceptable $p$-value is $\alpha$. 

The overall bootstrap based test procedure is (\cite{EffronTibshirani1994}, \cite{Silverman1997}, \cite{Golland_permutation_test}):
\begin{itemize}
\item
Repeat the following procedure \emph{B}$(=1000)$ times (with index $b = 1,\ldots,B$):
\begin{itemize}
\item
Generate resampled (with replacement) versions of the times series $X^N$, $Y^N$, denoted by $X_b^N$, $Y_{b}^N$ respectively.
\item
Compute the statistic $\theta^b$ = $\hat{I}(X_{b}^N \rightarrow Y_{b}^N)$.
\end{itemize}
\item
Construct an empirical CDF (cumulative distribution function) from these bootstrapped sample statistics, as
$F_\Theta(\theta) = P(\Theta \le \theta) = \frac{1}{B}\sum_{b=1}^B I_{x \ge 0}(x = \theta-\theta^b)$, where $I$ is an indicator random variable on its argument $x$.
\item
Compute the true detection statistic (on the original time series), $\theta_0 = \hat{I}(X^N \rightarrow Y^N)$ and its corresponding $p$-value ($p_0=1-F_\Theta(\theta_0)) $ under the empirical null distribution $F_\Theta(\theta)$.
\item
If $F_\Theta(\theta_0)  \ge (1-\alpha)$, then we have that the true DTI value is significant at level $\alpha$, leading to rejection of null-hypothesis (no directional association).
\end{itemize}

\subsubsection{Summary of DTI-based TF effector Inference}\label{summary}

Our proposed approach using DTI for determining the effectors for gene $B$ (\textit{Gata2} in the enhancer study) is as follows:
\begin{itemize}
\item
Identify the $G$ genes ($A_1,A_2,\ldots, A_G$), based on phylogenetic conservation (Fig. \ref{fig:TFBS_conservation}).
Preprocess the gene expression profiles by normalization and cubic spline interpolation. Assuming that there are $N$ points for each gene, entropy estimation is used to compute the terms in the DTI expression (Eqn. $2$).
\item For each pair of genes $A_i$ and $B$ among these $G$ genes : \\
\{
\begin{itemize}
\item
Look for a phylogenetically conserved binding site of TF encoded by gene $A_i$ in the upstream region of gene $B$.
\item
Find $DTI (A_i, B) = I(A_i^N \rightarrow B^N)$, and the normalized DTI from $A_i$ to $B$, $DTI(A_i,B) = \sqrt{1-e^{-2I(A_i^N \rightarrow B^N)}}$.
\item
Bootstrap resampling over the data points of $A_i$ and $B$ yields a null distribution for $DTI(A_i,B)$. If the true $DTI (A_i, B)$ is significant at level $\alpha$ with respect to this null histogram, infer a potential influence from $A_i$ to $B$.
\item
The value of the normalized DTI from $A_i$ to $B$ gives the putative strength of interaction/influence. \item Every gene $A_i$ which is potentially influencing $B$ is an `effector'. This search is done for each gene $A_i$ among these $G$ genes
($A_1,A_2,\ldots,A_G$).
\end{itemize}\}
\end{itemize}
\emph{Note}: As can be seen, phylogenetic information is inherently built
into the influence network inference step above. We note that, in this approach, the choice of potential effectors for a target gene is based on only those TFs that have a binding site at the target gene's promoter. This aims to reduce the overall search space based on biological prior knowledge.

As an example, we indicate the significance and strength of the DTI between the \textit{Pax2} TF and \textit{Gata2}. The high strength of influence and its significance coupled with the phylogenetic conservation of the \textit{Pax2} motif indicates expression evidence for the role of \textit{Pax2} in \textit{Gata2} regulation (\cite{GrimmondLittle2005}, \cite{pax2_gata2_pronephros}).
\begin{figure}[!t!h!b]
\centerline{\includegraphics[width=3in,height=1.6in]{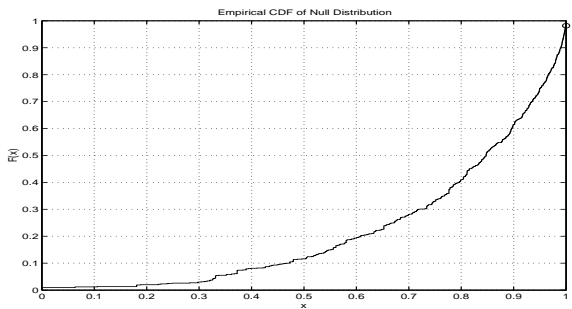}}
\caption{Cumulative Distribution Function for bootstrapped
$I(\textit{Pax2} \rightarrow \textit{Gata2})$ interaction. True $\hat{I}
(\textit{Pax2} \rightarrow \textit{Gata2}) = 0.9818$.
}\label{fig:pax2_gata2}
\end{figure}

\begin{figure}[!t!h!b]
\centerline{\includegraphics[width=3in,height=1.6in]{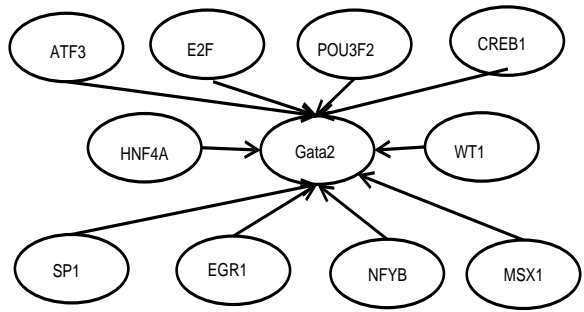}}
\caption{Putative upstream TFs using DTI for the \textit{Gata2}
gene.}\label{fig:effector_gata2}
\end{figure}

Such analysis can be extended to all TFs that are phylogenetically conserved.
For \textit{Gata2} regulation in the developing kidney, this set of putative TF effectors (apart from \textit{Pax2}) is shown in Fig. \ref{fig:effector_gata2}.

\subsection{Enhancer TF identification}\label{enh_TF}

In the earlier section, we have examined the identification of promoter TFs using phylogenetic sequence conservation of TFBS motifs in conjunction with expression level influence using DTI.  The next key step towards determining the nature of promoter-enhancer TF interactions is the identification of enhancer-TFs. As has been alluded to earlier, there is no method to precisely infer which transcription factors bind a certain regulatory element during long-range gene regulation. Thus, we appeal to a traditional approach of finding tissue-specific transcription factors that are phylogenetically conserved at any potential regulatory region (\cite{Enhancer_Prediction}, \cite{EnhancerBrowser2}). This is consistent with earlier observations that enhancers recruit tissue-specific transcription factors during the formation of the overall transcriptional machinery during gene expression, whereas promoters recruit components of the basal transcriptional machinery (\cite{Kleinjan2005}, \cite{Fraenkel2006}, \cite{Kreiman2004}, \cite{EnhancerBrowser2}, \cite{3C/4C}).

To ascertain the tissue-specificity of each TF that putatively binds a regulatory element (identified via phylogenetic conservation), we examine that TF's annotation in the UNIPROT database. This database is one of the most current sources of TF annotation and has details pertaining to the sequence specificity of the binding motif, the structure of the TF and its tissue-specificity of expression. For those TFs that do not have a UNIPROT annotation, we look at the tissue-expression of the corresponding gene from the mouse genome informatics (MGI) mRNA annotations. The MGI expression annotations encompass multiple modalities (literature, RNA \textit{in-situ}) to suggest a tissue-restricted or conversely, a ubiquitous expression of the TF gene. Thus, a set of tissue-specific transcription factors that bind any non-coding region of interest (such as an enhancer) can be identified (\cite{Fgf_enh}, \cite{EnhancerBrowser2}, \cite{Enhancer_Prediction}, \cite{Kreiman2004}, \cite{Fraenkel2006}). For the \textit{Gata2} UGEs, several potential TFs can be found, some of which are highlighted in Fig. \ref{fig:ppi_uge}.

\subsection{Enhancer-Promoter Distal Interaction via Protein-Protein Interactions - A Graph Based Analysis}\label{ppi_part2}

Using the notion of protein-protein interaction mediating long-distance interactions between promoters and enhancers during looping (\cite{looping_scan_track}, \cite{prom_enh_ppi1}, \cite{prom_enh_ppi2}), we explore the interactome to look for within-group and between-group interactions in the promoter-TF and the enhancer-TF groups.
The resultant interaction-graph can be examined for several ``structural" characteristics (like heterogeneity, degree distribution, path length, density, clustering coefficient and connected components) (\cite{netanalyzer_cytoscape}, \cite{Horvath}). The goal is to identify structural features that discriminate true enhancer vs. non-functional element activity based on their interaction-graph.

\begin{figure}[!t!h!b]
\centerline{\includegraphics[width=3in,height=1.6in]{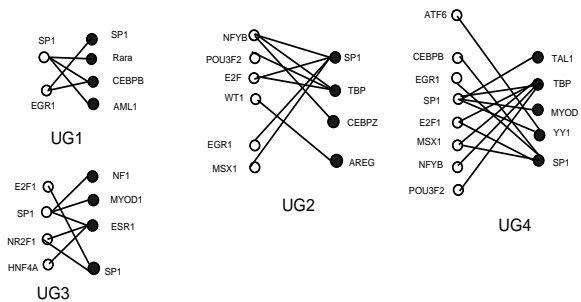}}
\caption{Protein-protein interaction between putative \textit{Gata2} TFs (hollow circles) and putative UG element TFs (filled circles). Note: This only shows the connections between two groups for one of the connected components. For our analysis, we consider both \textit{intra-} and \textit{inter-group} connections. From \emph{http://string.embl.de/}}\label{fig:ppi_uge}
\end{figure}

The interaction-graphs (e.g: Fig. \ref{fig:ppi_uge}) are obtained in the following manner:
\begin{itemize}
\item
One part of the graph (hollow circles) corresponds to the TF effector group at the promoter. These $V_p$ TFs are identified based on phylogenetic conservation and directed information (section: \ref{prom_TF}).
\item
The other part of the graph (filled circles) corresponds to the $V_e$ tissue-specific TFs group at the enhancer, identified based on phylogeny and annotation (section: \ref{enh_TF}).
\item
The interaction-graph is defined by the vertices $V = (V_p \cup V_e)$, and the edges $E = e_{i,j}$, $i,j \in (1,2, \ldots, |V_p\cup V_e|)$.
Each bidirectional edge $E=(e_{i,j})$ is derived from an annotated interaction between TFs $i$ and $j$, based on an interaction database. These edges describe both within-group TF interactions as well as between-group interactions. To obtain the TF interactions, we use protein-interaction information derived from the STRING (\emph{http://string.embl.de/}) and MiMI (\emph{http://mimi.ncibi.org/MiMI/home.jsp}) databases, both of which contain data derived from multiple sources, such as yeast-2-hybrid screens, literature, ChIP etc. Though there is some inherent noise in the accuracy of these high-throughput sources, they permit the use of a confidence threshold to discriminate a potentially true interaction from a spurious one. 
\end{itemize}

Though it would be of great value to use a catalog of gene-specific and tissue-specific regulatory regions (with all possible transcription factors) from which to find such interaction characteristics - such a repository does not yet exist.  In this section, we use a few examples (Gata3 OVE, Gata3 KE, Fgf OVE, Mecp2 F21/F6 , Shh FE) of known tissue-specific and gene-specific regulatory elements from literature, as a positive training set. For the negative training set, we consider the set of regions that were reportedly investigated in these transgenic experiments but did not yield gene-specific regulatory activity. Based on which structural metrics are associated with potential regulatory activity 
for these examples, we will examine if these features are predictive of \textit{Gata2} UGE enhancer behavior, from its interaction-graph.

We have presented a preliminary analysis of enhancer-promoter TF interaction-graphs for some genomic elements with known regulatory or non-regulatory activity (\cite{Mecp2}, \cite{Shh}, \cite{Gata3KE}, \cite{Fgf_enh}) in Table. \ref{ppi_properties}. The table represents the listing of the structural attributes of these interaction-graphs, following analysis methods from literature (\cite{mcode_cytoscape}, \cite{netanalyzer_cytoscape}, \cite{said_protein_network}). A brief summary of these attributes are given below. A deeper analysis of other graph topology metrics and their relation to functional enhancer activity is a topic of future interest.

\begin{table}[!h!b!t]
\centering
\begin{tabular}{lllllll}
Sequence & Class  &  Clustering  & Characteristic & Heterogeneity & Centralization & Density\\
         &        &  Coefficient & path length & & \\
Mecp2 F21\cite{Mecp2} & +1 & 0.208 &	2.824 &	0.668 &	0.184 &	0.133\\
Mecp2 F6 \cite{Mecp2}         & -1 & 0	& 1.75& 	0.342 &	0.067 &	0.145\\
Gata3 OVE \cite{Gata3KE}         & +1 & 0.036	 & 2.254 &	0.779	 & 0.359	& 0.154\\
Gata3 KE \cite{Gata3KE}    & +1 & 0.409	& 2.0	& 0.813	& 0.684	&0.216\\
Gata3 NE1 \cite{Gata3KE}         & -1 & 0.383	& 2.131	& 1.139	& 0.757	& 0.15 \\
Gata3 NE2 \cite{Gata3KE}         & -1 & 0.458 &	2.013 &	0.872 &	0.699 &	0.203\\
Fgf10 OVE \cite{Fgf_enh}        & +1 & 0.313 & 2.433 & 0.72 &	0.323 &	0.133 \\
Shh FE \cite{Shh} & +1 & 0.394 & 2.312	& 0.797	& 0.49	& 0.175
\\
\end{tabular}
\caption{The first column is the various regulatory and non-regulatory elements from literature, the next
column corresponds to its class label ($+1/-1$). The subsequent columns correspond to the attributes of the overall TF-interaction graph (both within-group and between-group interactions). }\label{ppi_properties}
\end{table}

\begin{itemize}
\item
Clustering coefficient: In undirected networks, the clustering coefficient $C_n$ of a node $n$ is defined as $C_n = 2e_n/[k_n(k_n-1)]$, where $k_n$ is the number of neighbors of $n$ and $e_n$ is the number of connected pairs between all neighbors of $n$. Thus $C_n$, of a node in a graph is the ratio of the number of edges between the neighbors of that node over the total number of edges that could exist among its neighbors. The clustering coefficient of a node is always a number between $0$ and $1$. The network clustering coefficient is the average of the clustering coefficients for all nodes in the network.
\item
Characteristic Path length: The length of a path along the graph is the number of hops (or edges) between any two nodes along the graph.  Though, there may be multiple paths between two nodes $n$ and $m$ (TFs) along the interaction-graph , the shortest path length $L(n,m) = (L(m,n))$ corresponds to the minimum across these multiple paths. This measure is computed for all pairs of nodes in the network. The characteristic path length denotes the average shortest-path distance of the graph. This gives the expected distance of any two connected nodes in the graph and is a global indicator of network-connectivity.
\item
Heterogeneity: Network heterogeneity denotes the coefficient of variation of the degree distribution. A network that is heterogeneous would consist of some nodes that are highly connected (exhibit `hub' behavior), while the majority of nodes tend to have very few connections. Understanding the heterogeneity of the degree distribution in biological networks is an interesting topic of current research, especially as a way to discover modularity \cite{Horvath}.
\item
Centralization: This refers to the overall connectivity (cohesion) of the graph. It indicates how strongly the graph is organized around its most central point(s). The central point(s) of the graph are the set of nodes which minimize the maximum distance distance from all other nodes in the graph \cite{SNA_book1}.
Networks whose topologies resemble a star/wheel pattern have a centralization close to $1$, whereas decentralized networks are characterized by having a centralization close to $0$. 
\item
Density: The neighborhood of a given node $n$ is the set of its neighbors. The connectivity of $n$, denoted by $k_n$, is the size of its neighborhood. The average number of neighbors indicates the average connectivity of a node in the network. A normalized version of this parameter is the network density $k_n/n(n-1)$. The density is a value between $0$ and $1$. It is also the average standardized degree. It shows how densely the network is populated with edges (i.e. how ``close-knit" an empirical graph is \cite{SNA_book1}, \cite{SNA_book2}). A network which contains no edges and solely isolated nodes has a density of $0$, whereas the density of a clique (completely connected graph) is $1$.
\end{itemize}

The above mentioned several network properties (as well as clustering coefficients, number of connected components etc.) are examined for the overall interaction-graphs for the reported enhancers from literature \cite{netanalyzer_cytoscape}.
A logistic regression as well as random forest analysis reveals that low values of heterogeneity, characteristic path length and centralization are fairly good predictors of potential enhancer activity. All of these attributes point to the decentralized, homogenous and somewhat tighter connectivity of the interaction-graphs for true enhancers. We note that the OOB error rate of the RF here is about $25\%$. The quality of this classifier can be expected to improve as we get more data from which to extract features.

We now examine the interaction-graphs for the test set, i.e. the four \textit{Gata2} UGEs. For illustration, we only show the largest connected component of the inter-group edges for each interaction graph (Fig. \ref{fig:ppi_uge}).

This figure indicates a very interesting property of the real enhancers vis-a-vis the other conserved elements. We see that the TF effectors for \textit{Gata2} such as \textit{SP1}, \textit{POU3F2} (identified in the TF effector network above, Fig. \ref{fig:effector_gata2}), are involved in cross-element interactions at the protein level, between the promoter and true enhancer ($UG2/4$). However, the network linkage in the elements that showed no enhancer activity is very sparse suggesting low cross-talk between promoter and enhancer. Also, the TFs at the enhancer nodes (dark circles) have a more uniform degree distribution in the functional elements $UG2/4$ as compared to the non-functional ones. Both these observations suggest lower heterogeneity and centralization of such functional interaction-graphs. Thus, the extent of TF cross-talk is a potential discriminator of possible enhancer function.  This shows that superimposing PPI information along with sequence and expression data helps reduce the number of false positives while integrating various aspects of distal regulation.

\section{Heterogeneous Data Integration and Validation on Gata2 UGEs}\label{data_integration}
As mentioned previously, the primary goal of the various methods developed above is to understand the behavior of known transcriptional elements along different genomic modalities. To validate their predictive potential, we have to demonstrate their application to predicting the behavior of the \textit{Gata2} UGEs (which is our test set). In this section, we present a framework that combines the results of the individual classifiers developed before (kidney-promoter RF, histone RF and interactome-RF) to obtain a integrated prediction. For combining heterogeneous classifiers, we will explore a ``probabilistic belief fusion" framework in this paper. Of course, other techniques from literature (like ensemble methods) are also highly amenable for exploration in this context.

The framework involves combining the `beliefs' of the individual classifiers to obtain a combined belief of prediction. To compute the belief of each classifier we start with examining the confusion matrices for each of the classifiers (kidney-promoter RF, histone-RF and graph-RF), following (\cite{BCC_Ghahramani}, \cite{classifier_combine1}, \cite{classifier_combine2}). Since each of the classifiers are random forests, we can obtain their OOB error estimates through these confusion matrices. For the graph-RF, this confusion matrix is as below,

\[
\mathbf{CM_{graph-RF}} =
\begin{pmatrix}
Class   &-1 & 1 & class.error \\
-1  & 4 & 1   & 0.20 \\
1   & 1 & 4   & 0.20
\end{pmatrix},\] thereby yielding an OOB error estimate of $\sim 20\%$.

Similarly, we have,
\[
\mathbf{CM_{promoter-RF}} =
\begin{pmatrix}
Class   &-1 & 1 & class.error \\
-1  & 67 & 19   & 0.22 \\
1   & 10 & 76   & 0.12
\end{pmatrix},\] thus yielding an OOB error estimate of $\sim 17\%$.

\[
\mathbf{CM_{histone-RF}} =
\begin{pmatrix}
Class   &-1 & 1 & class.error \\
-1  & 134 & 24   & 0.15 \\
1   & 21 & 119   & 0.15
\end{pmatrix},\] yielding an OOB error estimate of $\sim 15\%$.

\begin{figure}[!t!h!b]
\centerline{\includegraphics[width=3in,height=1.8in]{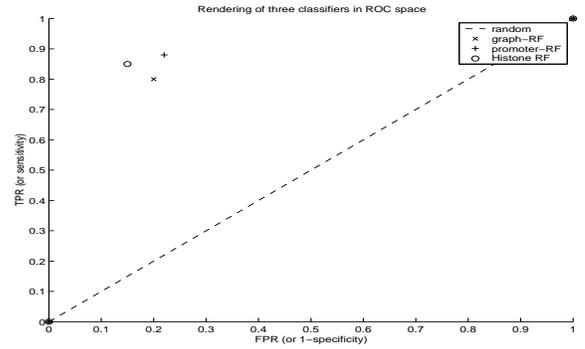}}
\caption{Representation of the three RF classifiers in ROC space (RF-promoter in $(+)$, and RF-histone in $(o)$, and graph-RF in $(\times)$). The diagonal line is the classification by random chance.}\label{fig:roc_rf}
\end{figure}

The three random forest classifiers are represented in ROC space (Fig. \ref{fig:roc_rf}). As can be seen, these three classifiers have fairly good performance characteristics.
Moreover these are three complementary data sources and can be effectively combined to improve detection reliability. Since they are trained on very different modalities, they can be assumed to be independent.

Each classifier is a function $e_k(x) = j_k$ that maps a data point ($x$) to the class $`j'$, with $k=1,2,\ldots,K$ and $j_k \in (-1,1)$. Here, $K = 3$, and $J=2$ classes.

Thus, the belief of the $k^{th}$ classifier is,
\begin{gather*}
bel_k(x \in C_i|e_k(x)=j_k) = P(x \in C_i|e_k(x)=j_k)
\end{gather*}
The overall belief, \textit{bel(i)}, given by,
\begin{multline*}
bel(i)= bel(x \in C_i|e_1(x)=j_1,\ldots,e_K(x)=j_K) =\\
P(x \in C_i|e_1(x)=j_1,\ldots,e_K(x)=j_K)\\
= \frac{P(e_1(x)=j_1,\ldots,e_K(x)=j_K|x \in C_i). P(x \in C_i)}{P(e_1(x)=j_1,\ldots,e_K(x)=j_K)}
\end{multline*}
Further, we have that,
\begin{gather*}
\frac{\prod_{k=1}^K P(e_k(x)=j_k|x \in C_i)}{\prod_{k=1}^K P(e_k(x) = j_k)} = \frac{\prod_{k=1}^K P(x \in C_i|e_k(x)=j_k)}{\prod_{k=1}^K P(x \in C_i)}
\end{gather*}
Thus,
\begin{gather*}
 bel(i) = P(x \in C_i). \frac{\prod_{k=1}^K P(x \in C_i|e_k(x)=j_k)}{\prod_{k=1}^K P(x \in C_i)}
\intertext{(due to independence of the $K$ classifiers,)}
\end{gather*}
In the absence of the posterior probability $P(x \in C_i)$, an approximation is used, leading to \cite{classifier_combine1},
\begin{gather*}
bel(C_i)= \frac{\prod_{k=1}^K P(x \in C_i|e_k(x)=j_k)}{\sum_{i=1}^J \prod_{k=1}^K P(x \in C_i|e_k(x)=j_k)}.
\end{gather*}
\textit{Note:} $J=2$ and $K=3$.
Depending on the belief value $bel(i)$, the decision rule $(E(x))$ for classifying data point $x$ is,
\begin{eqnarray*}
E(x) =j, \textrm{if \medspace} bel(j) = max_i \thinspace bel(i),\\
\textrm{or, } E(x)=j, \textrm{if \medspace} bel(j)= max_i \thinspace bel(i), \textrm{and, } bel(j) \ge \alpha,
\end{eqnarray*}where $0 < \alpha \le 1$, with $\alpha$ being a threshold.

\begin{table*}[!h!b!t]
\centering
\begin{tabular}{cccccc}
Sequence & True  &   Promoter RF & Histone RF & Interaction-graph RF & P(Class=+1)\\
         & Class &     prediction $e_1(x)$                  &   prediction $e_2(x)$           &   prediction $e_3(x)$               & (Overall Belief)\\
\textit{Gata2} UG1 & -1 & -1 & -1 & -1 & 0.0054 \\
\textit{Gata2} UG2 & +1 & +1 & +1& +1 & 0.9875 \\
\textit{Gata2} UG3 & -1 & +1 & +1 & -1 & 0.832 \\
\textit{Gata2} UG4 & +1 & +1 & +1& +1 & 0.9875 \\
\end{tabular}
\caption{Combined belief generation during heterogeneous classifier integration. The last column represents the combined belief (probability that the UG sequence is an enhancer) as a result of integrating the promoter-RF, histone-RF and graph-RF predictions.}\label{class_integration}
\end{table*}

We now show the output classes of each of the $3$ classifiers as well as the combined belief on the \textit{Gata2} UGEs in Table. \ref{class_integration}. More specifically, for the first row in Table. \ref{class_integration}, the overall belief equation above becomes,

\begin{multline*}
bel(ug1=+1)=
\frac{P(ug1=+1|e_1(x)=-1).P(ug1=+1|e_2(x)=-1).P(ug1=+1|e_3(x)=-1)}
{\splitfrac{P(ug1=+1|e_1(x)=-1).P(ug1=+1|e_2(x)=-1).P(ug1=+1|e_3(x)=-1)+}
{P(ug1=-1|e_1(x)=-1).P(ug1=-1|e_2(x)=-1).P(ug1=-1|e_3(x)=-1)}}
= 
\\ \frac{[(1-prec_{n,1}) \times (1-prec_{n,2})] \times [(1-prec_{n,3})}{(1-prec_{n,1}) \times (1-prec_{n,2}) \times (1-prec_{n,3})]+[prec_{n,1} \times prec_{n,2} \times prec_{n,3}]}
\end{multline*}
Here, $prec_{n,k} = \frac{TN_k}{TN_k+FN_k}$. Similarly, $prec_{p,k}=\frac{TP_k}{TP_k+FP_k}$. These are the negative and positive precision values respectively, for the $k^{th}$ classifier. These rates are obtained from the corresponding confusion matrices shown above. This approach is followed for each of the $UG1-4$ elements.

If we set a threshold of $\alpha=0.85$ or $0.90$, we would get $UG2$ and $UG4$ to be the true enhancers ($100\%$ accuracy). However, for a choice of $\alpha=0.8$, $UG3$ is predicted to be an enhancer in spite of being declared a member of the $(-1)$ class by the graph-RF. This choice of threshold thus determines the performance of the combined classifier.

Under the $\alpha=0.8$ case, however, the results are not to be interpreted as a $25\%$ error rate since the nature of the test set (\textit{Gata2} UG enhancers) are very different from the training data of each modality (promoters are proximal elements whereas enhancers are distal; histone sequences are for a different cell-context; and interaction-graphs are obtained over different genes). The fact that we are getting such good prediction in spite of the training sets being so different is a strong point in favor of examining and integrating these data sources. The real test-error rates are given by the OOB error estimates of the individual classifiers. 

\section{Summary of Approach} \label{summary}
In this work, we  have shown that,
\begin{itemize}
\item
Motif signatures are predictive of regulatory element location. These comprise sequence-motifs derived from tissue-specific gene promoter sequences as well as sequences related to epigenetic preferences during gene regulation.
\item
Promoter and enhancer TFs that are putatively recruited during gene (\textit{Gata2}) regulation can be identified using a combination of phylogenetic conservation, expression data, and tissue-specificity annotation.
\item
Effector TFs (via DTI) at the gene proximal promoter have high network linkage with enhancer TFs in case of functional enhancers. The TF interaction-graphs of truly functional elements are seen to be have a lower centralization, characteristic path length and heterogeneity suggesting higher cross-talk during formation of the transcription factor complex.
\end{itemize}

These perspectives (based on sequence, expression and interactome data) shed some light on the sequence and mechanistic preferences of true regulatory regions interspersed genome-wide. It is to be noted that this model is data driven and may not directly correspond to the biology of transcription. However, much like markov models for gene sequence annotation, we believe that such data-driven models are useful for model-building during genome-wide study.


\section{Conclusions}\label{conclusions}


In this work, we have examined the problem of regulatory element identification. Such an effort has implications to understand the genomic basis of key biological processes such as  development and disease. Using the biophysics of transcription, this can be modeled as a problem in data integration over various experimental modalities such as sequence, expression, transcription factor binding and interactome-data. Using the case study of enhancers corresponding to the \textit{Gata2} gene, we examine the utility of these heterogeneous data sources for predictive feature selection, using principled methodologies and metrics.

Based on motif signatures, we find that they predict the true enhancers ($UG2$, $UG4$), and the false enhancer $UG1$, but mispredict $UG3$ to be an enhancer. However, a  mechanistic insight that analyzes enhancer behavior based on the interactions between distally and proximally recruited transcription factors can greatly improve on prediction accuracy. Additionally, combining heterogeneous classifiers based on multiple data modalities yields an improved accuracy of prediction.

The novelty of the proposed work spans several areas. Firstly, data sources that are relevant to understand the mechanism of gene regulation (with \textit{Gata2} as an example) have been identified. We have developed methods that reconcile the behavior of known regulatory elements along each of these modalities. The kidney-promoter based classifier aims to discover sequence preferences of kidney-specific regulatory regions. The utilization of histone-modified sequences and their exploration for sequence motifs are indicative of epigenetic-preferences and nucleosome-occupancy patterns. This has not been explored before in the realm of LRE characterization. The use of DTI as a metric to infer putative TF to target-gene influence is a recent one that serves to integrate phylogenetic TFBS conservation along with expression data. Finally, the utilization of graph-based analysis techniques to understand the ``structure" of the TF interaction-graph between enhancer and promoter helps us understand true enhancer behavior from a mechanistic viewpoint. The probabilistic combination of multiple classifiers (each deriving from a unique data resource) aims to reconcile the behavior of existing enhancers along multiple modalities. We hope to demonstrate that a principled integration of non-overlapping genomic modalities can be used to interpret the context and specificity of gene regulation.

\section{Future Work}

Some key elements directly emerge for guiding future research. As already alluded to in the motif-signature procedure, specific expression data corresponding to stages and tissues of interest would greatly improve the specificity of regulatory element prediction.  Furthermore, as histone modification maps for different cell lines  are generated, the false positive rate of prediction would decrease, thereby improving accuracy. Several other learning paradigms can be introduced into this setting, since we are learning from structured data. Also, methods in joint classifier and feature optimization might likely improve the accuracy of predictions. Additionally, methods that analyse the grammar of these cis-regulatory regions (LREs) and look for motif position, spacing  and orientation will be of great utility. 

At the expression level, methods for supervised network inference would have a great impact on the discovery of TF effectors. Rapid advances have been made in this area and their relevance to the biological context of the problem has become very principled. At the interactome level, the work presented here can be extended to the investigation of graph-clusters for weighted interaction-graphs. The weighted edges are obtained from the confidence of the individual data sources, as well as the number of species over which that particular edge is conserved (\cite{mcode_cytoscape}, \cite{ideker_network}). Such analysis enables the discovery of subgraphs of various degrees of inter-connectedness, thereby discovering functional ``graph-motifs".

\section{Acknowledgements}
The authors gratefully acknowledge the support of the NIH under award 5R01-GM028896-21 (J.D.E). We would like to thank Prof. Sandeep Pradhan, Mr. Ramji Venkataramanan and Mr. Dinesh Krithivasan for useful discussions on directed information. Ms. Swapnaa Jayaraman at the University of Michigan is gratefully acknowledged for discussions about network-attributes. We are grateful to Prof. Erik Learned-Miller and Dr. Damian Fermin for sharing their code for
high-dimensional entropy estimation and ENSEMBL sequence extraction, respectively. We are very grateful to the anonymous reviewers and the associate editor for their constructive comments and feedback to improve the quality of the manuscript.

\section{Availability}
The source code of the analysis tools (in R $2.0$ and MATLAB $6.1$)
is available on request. Supplementary data (the set of various tissue-specific promoters, enhancers, housekeeping promoters and hexamer lists for each category) will be made available from the publisher's website.


\end{document}